\documentclass[american,twocolumn,aps,pre]{revtex4-1}
\usepackage[T1]{fontenc}
\usepackage[latin9]{inputenc}
\usepackage{color}
\usepackage{mathrsfs}
\usepackage{amsbsy}
\usepackage{amstext}
\usepackage{amssymb}
\usepackage{graphicx}
\usepackage{esint}

\makeatletter

\providecommand{\tabularnewline}{\\}

\makeatother

\usepackage{babel}
\begin{document}

\title{Bending and Gaussian rigidities of confined soft spheres from
second-order virial series}

\author{$^{*\dagger}$Ignacio Urrutia}

\email{iurrutia@cnea.gov.ar}

\selectlanguage{american}%

\affiliation{$^{*}$Departamento de Física de la Materia Condensada, Centro Atómico
Constituyentes, CNEA, Av.Gral.~Paz 1499, 1650 Pcia.~de Buenos Aires,
Argentina}

\affiliation{$^{\dagger}$CONICET, Avenida Rivadavia 1917, C1033AAJ Buenos Aires,
Argentina}
\begin{abstract}
We use virial series to study the equilibrium properties of confined
soft-spheres fluids interacting through the inverse-power potentials.
The confinement is induced by hard walls with planar, spherical and
cylindrical shapes. We evaluate analytically the coefficients of order
two in density of the wall-fluid surface tension $\gamma$ and analyze
the curvature contributions to the free energy. Emphasis is in bending
and Gaussian rigidities, which are found analytically at order two
in density. Their contribution to $\gamma(R)$ and the accuracy of
different truncation procedures to the low curvature expansion are
discussed. Finally, several universal relations that apply to
low-density fluids are analyzed. 
\end{abstract}
\maketitle

\section{Introduction\label{sec:Intro}}

Inhomogeneous fluid systems with interfaces have been studied for
a long time and are ubiquitous in nature. Characteristic examples
of such systems are the two phase coexistence with vapor-liquid interface
and the confined system with fluid-wall interface. In the second case
the interface is induced by an external potential that yields spatial
regions forbidden for the fluid. From a thermodynamic perspective
the correspondence between the free-energy of the system and the shape
of its interface is a relevant topic both for basic and applied investigation.

Confined fluids enable us to study in a simple manner the dependence
of the interface free energy with the interface shape by simply changing
the shape of the vessel. In particular, smooth interfaces are appropriate
to analyze the deviation from the well-known planar limit where the
theoretical framework is established. Even for low-density confined
fluids the first principles theories based in virial series approach
are still under development. Seminal work of Bellemans dates from the
1960's \citep{Bellemans_1962,Bellemans_1962_b,Bellemans_1963} and
later developments of Rowlinson and McQuarrie \citep{Rowlinson_1985,McQuarrie_1987}
were done in the 1980's. Recently, new exact results based on virial
series were obtained for confined hard spheres (HS),\citep{Yang_2013,Urrutia_2014,Urrutia_2014b}
square well, and even Lennard-Jones, systems.\citep{Urrutia_2016}
This work aims to contribute in this direction by studying the physical
properties of pure repulsive soft-spheres system confined by curved
walls.

The soft-sphere particles interact through a tuneable softness core
(without an attractive well) produced by the inverse-power law potential
(IPL). This model has interesting scaling properties\citep{Hoover_1970,Rosenfeld_1983,Kohl_2014}
and constitutes an important reference to study more complex systems.\citep{Hansen_1970,Kambayashi_1994,Hummel_2015}
Several studies focused on elucidating the relation between core-softness
and thermodynamic properties.\citep{Lange_2009,Shi_2011,Wheatley_2013,Zhou_2013b}
Basic research about bulk transport and virial coefficients was started
by Rainwater and others,\citep{Rainwater_1978,Rainwater_1979,Dixon_1979,Kayser_1980,Rainwater_1981,Rainwater_1984}
and continues up to present.\citep{Wheatley_2005,Tan_2011,Barlow_2012,Wheatley_2013}
Analytic equations of state of the soft-sphere fluid were found using
as input the known bulk virial coefficients using resummation, by
adapting the Carnahan Starling equation of state for HS to soft-spheres
and utilizing Padé approximants.\citep{Maeso_1993,Tan_2011,Barlow_2012}
Aspects of recent research interest in the soft-sphere system are
the scaling law invariance of its properties,\citep{Pieprzyk_2014,Hummel_2015}
the enhancement of effective attraction between colloids produced
by the soft repulsion in colloid+depletants system,\citep{Rovigatti_2015}
the equilibrium and nonequilibrium dynamics of particles,\citep{Ding_2015}
and the analysis of the sound velocity near the fluid-solid phase
transition.\citep{Khrapak_2016} 

We will study the dependence on curvature of equilibrium thermodynamic
properties of the fluid confined by curved walls based on its \emph{inhomogeneous}
second virial coefficient. For simplicity only constant-curvature
surfaces, i.e., planar, spherical, and cylindrical, are considered. The
expansion of the wall-fluid surface tension on the surface curvature
follows the Helfrichs expression.\citep{Helfrich_1973} Applied to
the sphere and cylinder symmetry the expansion of $\gamma(R)$ gives
\begin{eqnarray}
\gamma_{\textrm{s}}(R) & = & \gamma-\frac{2\gamma\delta}{R}+\frac{2k+\bar{k}}{R^{2}}+\frac{\mathscr{C}}{R^{3}}+\ldots\:,\label{eq:gammaks}\\
\gamma_{\textrm{c}}(R) & = & \gamma-\frac{\gamma\delta}{R}+\frac{k}{2R^{2}}+\ldots\:,\label{eq:gammakc}
\end{eqnarray}
where dots represent higher-order terms in $R^{-1}$. Here, $\gamma$
is the wall-fluid surface tension for a planar surface and $\delta$
is the (radius-independent) Tolman length, which is related with the
total curvature. Next term beyond $\gamma\delta$ includes the bending
rigidity $k$ (associated with the square of the total curvature)
and the Gaussian rigidity $\bar{k}$ (associated with Gaussian curvature).
In the present work we will analyze Eqs. (\ref{eq:gammaks},\ref{eq:gammakc})
using virial series expansion. 

In the following Sec. \ref{sec:Theory} it is given a brief review
of the statistical mechanics virial series approach to inhomogeneous
fluids. The second-order cluster integral is analytically evaluated
for the confined soft-sphere system interacting through IPL in Sec.
\ref{sec:SndOrderTerms}. There, the functional dependence on the
hardness parameter $\nu$, the temperature, and the radius is shown.
Surface tension is studied at low density as a function of $\nu$
and $R$ in Sec. \ref{sec:Results}. In Sec. \ref{sec:Rigidities}
the bending and Gaussian curvature rigidity constants are extracted
and studied as a function of $\nu$. It is found that for $\nu=6$
there exists a logarithmic term in the surface free-energy that corresponds
to curvature rigidities and that is absent for $\nu>6$. Several
recently found universal relations that apply to any fluid are here
verified for soft spheres. Besides, our exact results and some of
these universal relations are used to test the degree of accuracy
of morphometric approach at low density. Finally, a summary is given
in Sec. \ref{sec:Summary}.

\section{Statistical Thermodynamic background\label{sec:Theory}}

The following short summary about virial series for confined systems
attempts to give a closed-form of the general theory and contains
a collection of ideas and formulae taken from Refs. \citep{Rowlinson_1985,Urrutia_2011_b,Urrutia_2012,Yang_2013}.
The virial series of the free energy here developed will be used in
Secs. \ref{sec:SndOrderTerms} to \ref{sec:Rigidities} to study the
confined IPL system up to order two in the activity (the lowest nontrivial
order).

We consider an inhomogeneous fluid at a given temperature $T$ and
chemical potential $\mu$ under the action of an external potential.
The grand canonical ensemble partition function (GCE) of this system
is 
\begin{equation}
\Xi=1+\sum_{n=1}\lambda^{n}Q_{n}\:,\label{eq:gcO}
\end{equation}
where $\lambda=\exp(\beta\mu)$ and $\beta=1/k_{\textrm{B}}T$ is
the inverse temperature ($k_{\textrm{B}}$ is the Boltzmann's constant).
In Eq. (\ref{eq:gcO}) $Q_{n}$ is the canonical ensemble partition
function
\begin{eqnarray}
Q_{n} & = & \Lambda^{dn}Z_{n}/n!\:,\label{eq:Qn-1}\\
Z_{n} & = & \int g_{n}\left(\mathbf{x}\right)\exp\left(-\beta\phi_{(n)}\right)d\mathbf{x}\:,\label{eq:Zn-1}
\end{eqnarray}
where $\Lambda$ is the de Broglie thermal wavelength and $d$ is
dimension. $Z_{n}$ is the configuration integral, $\phi_{(n)}$ is
the interaction potential between particles, $g_{n}\left(\mathbf{x}\right)=\prod_{i=1}^{n}g\left(\mathbf{x}_{i}\right)$,
$g\left(\mathbf{x}_{i}\right)=\exp\left(-\beta\psi_{i}\right)$, and
$\psi_{i}$ is the external potential over the particle $i$.

In Eq. (\ref{eq:gcO}) the sum index may end either at a given value
representing the maximum number of particles in the open system or
at infinity. Fixing this value one may study small systems.\citep{Rowlinson_1986}
The main link between GCE and thermodynamics is still through the
grand free energy $\Omega$,
\begin{equation}
\beta\Omega=-\ln\Xi\:.\label{eq:OmgLog}
\end{equation}
Some thermodynamic quantities could be directly derived from $\Omega$
as, e.g., the mean number of particles $\left\langle n\right\rangle =-\beta\lambda\partial\Omega/\partial\lambda$.
Yet, other quantities could be derived from $\Omega$ once volume
and area measures of the system are introduced. For fluids confined
in regions of volume $V$ bounded by constant curvature surfaces with
area $A$ the grand free energy can be decomposed as 
\begin{equation}
\Omega=-PV+\gamma A\:,\label{eq:OmegaGam}
\end{equation}
with bulk pressure $P=-\left.\frac{\partial\Omega}{\partial V}\right|_{\mu,T,A,R}$
and fluid-substrate surface tension $\gamma=\left(\Omega+PV\right)/A$.

In the GCE, several quantities can be expressed as power series in
the activity $z=\lambda/\Lambda^{3}$ (virial series in $z$), with
cluster integrals $\tau_{j}$ as coefficients. The most frequent
in the literature are 
\begin{eqnarray}
\beta\Omega & = & -\sum_{j=1}^{\infty}\frac{z^{j}}{j!}\tau_{j}\:,\label{eq:Omgz}\\
\left\langle n\right\rangle  & = & \sum_{j=1}^{\infty}\frac{jz^{j}}{j!}\tau_{j}\:.\label{eq:meanN}
\end{eqnarray}
For inhomogeneous fluids it is convenient to define the $n$-particles
cluster integral $\tau_{n}$ as
\begin{equation}
\tau_{n}=n!\int g_{n}(\mathbf{x})\,b_{n}\left(\mathbf{x}_{1},\ldots,\mathbf{x}_{n}\right)d\mathbf{x}\:,\label{eq:Taun}
\end{equation}
where $b_{n}\left(\mathbf{x}_{1},\ldots,\mathbf{x}_{n}\right)$ is
the Mayer's cluster integrand of order $n$. To obtain Eq. (\ref{eq:Omgz})
from Eqs. (\ref{eq:gcO}) and (\ref{eq:OmgLog}) we follow the regular
diagrammatic expansion.\citep{Hansen2006} For homogeneous systems
$g(\mathbf{x})=1$ in Eq. (\ref{eq:Taun}) and therefore $b_{n}(\mathbf{x})$
does not depend on the position of the cluster producing the usual
Mayer cluster coefficient $b_{n}$. Thus, performing an extra integration
\begin{equation}
\tau_{n}=n!\int_{\infty}b_{n}(\mathbf{r})d\mathbf{r}=n!Vb_{n}\:,\label{eq:Taunh}
\end{equation}
with $V$ the volume of the accessible region, i.e., the infinite
space or the cell when periodic boundary conditions are used.\citep{Hill1956}
Eqs. (\ref{eq:OmegaGam},\ref{eq:Omgz},\ref{eq:Taunh}) give the
pressure virial series in powers of $z$ for the bulk system and using
Eq. (\ref{eq:meanN}) the standard virial series for $\beta P$ in
power of number density can be obtained.

\section{Evaluation of second cluster integral\label{sec:SndOrderTerms}}

We focus on the case of an external potential $\psi$, which is zero
if $\mathbf{r}\in\mathcal{A}$ and infinite otherwise. Furthermore,
$\partial\mathcal{A}$ (the boundary of $\mathcal{A}$) is a surface
with constant curvature characterized by an inverse radius $R^{-1}$
for spherical or cylindrical surfaces, that is zero in the planar
case. Therefore $Z_{1}$, the CI of one-particle system, coincides
with $V$, the volume of $\mathcal{A}$ and $A$ corresponds with
the boundary area. Thus, $\tau_{1}=V$, which is enough to describe
the confined ideal gas.

The first nontrivial cluster term is that of second order. It describes
the physical behavior of the inhomogeneous low-density gases up to
order two in $z$. We consider a system of particles interacting through
a spherically symmetric pair potential $\phi(r)$ with $r=\bigl|\mathbf{r}_{2}-\mathbf{r}_{1}\bigr|$
the distance between particles. For the second-order cluster we have
$b_{2}(r)=f(r)$ in terms of the Mayer's function $f(r)=\exp\left(-\beta\phi\right)-1$.
To evaluate $\tau_{2}$ we adapt and simplify here the approach followed
in Ref. \citep{Urrutia_2016}. Introducing the identity $g\left(\mathbf{x}_{1}\right)g\left(\mathbf{x}_{2}\right)=g\left(\mathbf{x}_{1}\right)-g\left(\mathbf{x}_{1}\right)\left[1-g\left(\mathbf{x}_{2}\right)\right]$
in Eq. (\ref{eq:Taun}) and rearranging terms $\tau_{2}$ reads
\begin{eqnarray}
\tau_{2} & = & 2Z_{1}b_{2}+\Delta\tau_{2}\:,\label{eq:Tau2}
\end{eqnarray}
with $b_{2}$ the second cluster integral for the bulk system and
\begin{eqnarray}
\Delta\tau_{2} & = & -\int_{0}^{r_{max}}f(r)w(r)dr\:,\label{eq:DTau2}\\
w(r) & = & \int_{u_{min}(r)}^{u_{max}(r)}S(u)s(r,u)du\:.
\end{eqnarray}
Here $u$ is the distance between particle one and $\partial\mathcal{A}$,
$S(u)$ is the area of the surface parallel to $\partial\mathcal{A}$
that lies in $\mathcal{A}$ at a distance $u$ and $s(r,u)$ is the
surface area of a spherical shell with radius $r$ (with the center
in $\mathcal{A}$ at distance $u$ from $\partial\mathcal{A}$) that
lies \emph{outside} of $\mathcal{A}$. By definition function $w(r)$
is thus purely geometric. A representation of $S(u)$ and $s(r,u)$
can be seen in Fig. 1 of Ref. \citep{Urrutia_2016}. Further, one
finds 
\begin{eqnarray}
2Z_{1}b_{2} & = & \int_{0}^{r_{max}}f(r)W(r)dr\:,\label{eq:Tau2b}\\
W(r) & = & s(r)\int_{0}^{u_{max}}S(u)du=s(r)V\:,
\end{eqnarray}
being $s(r)=4\pi r^{2}$ (the surface of the sphere with radius $r$).
Eqs. (\ref{eq:Tau2},\ref{eq:DTau2},\ref{eq:Tau2b}) give
\begin{eqnarray}
\tau_{2} & = & \int_{0}^{r_{max}}f(r)\bar{w}(r)dr\:,\label{eq:Tau2Inho}\\
\bar{w}(r) & = & \int_{0}^{u_{max}}S(u)\,\bar{s}(r,u)du\:,
\end{eqnarray}
with $\bar{w}(r)=W(r)-w(r)$ and being $\bar{s}(r,u)=s(r)-s(r,u)$
the surface area of a spherical shell of radius $r$ (with the center
in $\mathcal{A}$ at distance $u$ from $\partial\mathcal{A}$) that
lies \emph{inside} of $\mathcal{A}$. Eq. (\ref{eq:DTau2}) was derived
previously in Ref. \citep{Urrutia_2016} where it was used to evaluate
$\Delta\tau_{2}$ for the confined Lennard Jones system. On the other
hand, Eq. (\ref{eq:Tau2Inho}) is new and will be used in present
work to directly solve $\tau_{2}$ without intermediate steps.

When $\partial\mathcal{A}$ is a planar or spherical surface, $w(r)$
and $\bar{w}(r)$ are polynomial in $r$, while for cylindrical surfaces
both functions can be approximated for large radii as a truncated
series in $R^{-1}$, which gives a polynomial in $r$ too.\citep{Urrutia_2016}
Note that $b_{2}$ in Eq. (\ref{eq:Tau2b}) involves the dependence
$W(r)\propto r^{2}$ showing that if the bulk system is analytically
tractable then $\tau_{2}$ of the confined system {[}in Eq. (\ref{eq:Tau2Inho}){]}
would also be. Thus, Eq. (\ref{eq:Tau2Inho}) is a good starting point
to evaluate $\tau_{2}$ for systems of particles confined by a single
surface with spherical, cylindrical or planar shape.

We introduce the IPL pair interaction,
\begin{equation}
\phi(r)=\alpha\left(\frac{r}{\sigma}\right)^{-\nu}\:,\label{eq:POWpot}
\end{equation}
with $\alpha>0$ and being $\nu$ the hardness parameter. This fixes
$f(r)$ in Eq. (\ref{eq:Tau2Inho}). The case $\nu=12$ is used to
model pure repulsive molecules, yet higher values like $\nu=18,$
or $36$ are utilized in studies of short-range repulsive macroscopic
particles as is the case of neutral colloids and colloid-depletant
interaction.\citep{Vliegenthart_2000,Rovigatti_2015} To obtain $\tau_{2}$
from Eq. (\ref{eq:Tau2Inho}) we shall solve integrals of the type
\begin{equation}
C_{m+1,k}=\int_{0}^{l}\left[\exp\left(-\tilde{\beta}x^{-\nu}\right)-1\right]x^{m}dx\:,\label{eq:Cx}
\end{equation}
where $x=r/\sigma$, $\tilde{\beta}=\beta\alpha$ is an adimensional
inverse temperature and $l$ is typically $2R/\sigma$ or $\infty$.
Changing variables to $x^{\nu}$ we obtain $C_{m+1,\nu}=\frac{1}{\nu}C_{q,1}$
where $q=\frac{m+1}{\nu}$ (also, $l$ in $C_{m+1,\nu}$ is replaced
by $l^{\nu}$ in $C_{q,1}$). Changing variables again we found
\begin{eqnarray}
C_{q}(\varepsilon) & = & \int_{\varepsilon}^{\infty}y^{-(1+q)}\left[\exp\left(-\tilde{\beta}y\right)-1\right]dy\:.\label{eq:Cy}\\
 & = & \tilde{\beta}^{q}\Gamma(-q,\tilde{\beta}\varepsilon)-\frac{\varepsilon^{-q}}{q}
\end{eqnarray}
where $\varepsilon=l^{-\nu}$, $\Gamma(a,x)$ is the incomplete gamma
function\citep{Abramowitz1972}, and $C_{q,1}$ was replaced by $C_{q}(\varepsilon)$.
In Appendix \ref{Apsec:Gamma} we resume the relevant properties of
$C_{q}$ including its behavior at $0<\varepsilon\ll1$ and $\varepsilon\gg1$.
An alternative to the potential given in Eq. (\ref{eq:POWpot}) is
the inclusion of a short range hard-core repulsion. For completion,
the function $C_{q}$ for this pair interaction is given in Appendix
\ref{Apsec:HardCore}.

In terms of $C_{q}(\varepsilon)$ the result for the bulk is $\frac{\tau_{2}}{2}=V\frac{2\pi}{\nu}C_{3/\nu}(0)$.
We obtain the following expressions of $\tau_{2}$ for the confined
fluid:
\begin{equation}
\frac{\tau_{2}}{2}=V\frac{2\pi}{\nu}C_{3/\nu}(0)-A\frac{\pi}{2\nu}C_{4/\nu}(0)\:,\label{eq:Tau2plan}
\end{equation}
\begin{equation}
\frac{\tau_{2}}{2}=V\frac{2\pi}{\nu}C_{3/\nu}(\varepsilon)-A\frac{\pi}{2\nu}C_{4/\nu}(\varepsilon)+\frac{\pi^{2}}{6\nu}C_{6/\nu}(\varepsilon)\:,\label{eq:Tau2sphe}
\end{equation}
\begin{eqnarray}
\frac{\tau_{2}}{2} & = & V\frac{2\pi}{\nu}C_{3/\nu}(\varepsilon)-A\frac{\pi}{2\nu}C_{4/\nu}(\varepsilon)+\nonumber \\
 &  & \frac{L}{R}\frac{\pi^{2}}{32\nu}C_{6/\nu}(\varepsilon)+\frac{L}{R^{3}}\frac{\pi^{2}}{1024\nu}C_{8/\nu}(\varepsilon)+....\:,\label{eq:Tau2cyl}
\end{eqnarray}
where Eq. (\ref{eq:Tau2plan}) applies to the planar case and Eqs.
(\ref{eq:Tau2sphe},\ref{eq:Tau2cyl}) correspond to the confinement
in spherical and cylindrical cavities, respectively. In Eqs. (\ref{eq:Tau2sphe},\ref{eq:Tau2cyl})
and from now on we fix $\varepsilon=\left(2R\right)^{-\nu}$ ($\sigma$
is the unit length). For the cylindrical case, higher-order $C_{q}(\varepsilon)$
functions were omitted. Truncation of Eq. (\ref{eq:Tau2cyl}) produces
an spurious term proportional to $R^{5}$ that should be removed {[}e.g.,
if we discard terms beyond $C_{8/\nu}(\varepsilon)$ one must compensate
Eq. (\ref{eq:Tau2cyl}) with the addition of a term $-\frac{163}{96}\pi^{2}LR^{5}${]}.
$\tau_{2}$ for systems outside of sphere or cylinder follows directly
from Eqs. (\ref{eq:Tau2sphe},\ref{eq:Tau2cyl}) and Eq. (\ref{eq:Tau2}),
considering that $\Delta\tau_{2}$ remains unmodified. In the
limit of large $R$ (i.e. $\varepsilon\rightarrow0$) the behavior
of $C_{q}(\varepsilon)$ is as follows: if $0<q<1$ then $C_{q}(\varepsilon)\approx\tilde{\beta}^{q}\Gamma(-q)+\tilde{\beta}\,\varepsilon^{1-q}$
being $\Gamma(-q)<0$, if $q=1$ then $C_{q}(\varepsilon)\approx\tilde{\beta}\ln(\tilde{\beta}\varepsilon)$,
and if $q>1$ (and non-integer values) then $C_{q}(\varepsilon)\approx\tilde{\beta}\,\varepsilon^{1-q}$.
Expressions (\ref{eq:Tau2plan},\ref{eq:Tau2sphe},\ref{eq:Tau2cyl})
are formally identical to that obtained previously for different types
of pair potentials which produce a different expression for $C_{q}(\varepsilon)$.\citep{Urrutia_2016}

For short-range potentials, those with $\nu>6$, we found
\begin{equation}
\frac{\tau_{2}}{2}=Vb_{2}-Aa_{2}+s\left[c_{2}+\mathcal{O}(R^{6-k})\right]\:,\label{eq:Tau2short}
\end{equation}
where coefficients $b_{2}$, $a_{2}$ and $c_{2}$ are
\begin{eqnarray}
b_{2} & = & -\frac{2\pi}{3}\tilde{\beta}^{3/\nu}\Gamma\left(1-\frac{3}{\nu}\right)\:,\label{eq:b2}\\
a_{2} & = & -\frac{\pi}{8}\tilde{\beta}^{4/\nu}\Gamma\left(1-\frac{4}{\nu}\right)\:,\label{eq:a2}\\
c_{2} & = & -\frac{\pi^{2}}{36}\tilde{\beta}^{6/\nu}\Gamma\left(1-\frac{6}{\nu}\right)\:,\label{eq:c2}
\end{eqnarray}
and we have defined $s=0$ for planar, $s=1$ for spherical, and $s=\frac{3L}{16R}$
for cylindrical surfaces. Eq. (\ref{eq:b2}) is consistent with the
known analytic expression for the second bulk virial coefficient.\citep{Barlow_2012}
One notes that in Eq. (\ref{eq:Tau2short}) $s\times c_{2}$ term
scales with $A/R^{2}$ while a term scaling with $A/R$ is absent.
For spherical walls we also calculate the term of order $R^{-1}$,
which is $d_{2}/R$ with $d_{2}=\frac{\pi^{2}\tilde{\beta}}{24}$
if $\nu=7$ and $d_{2}=0$ if $\nu>7$. At $\nu\rightarrow\infty$
IPL potentials behave as those of HS. To analyze the deviation from
the HS behavior we obtained the asymptotic hardness expansion\citep{Rainwater_1978,Rainwater_1979}
with $\Gamma\left(1-\frac{m+1}{\nu}\right)\approx1+\gamma_{\textrm{E}}\frac{m+1}{\nu}$
and being $\gamma_{\textrm{E}}\approx0.57721$ the Euler constant.
For non short-range potentials Eq. (\ref{eq:Tau2short}) must be modified.
For $\nu=6$ we found
\begin{equation}
\frac{\tau_{2}}{2}=Vb_{2}-Aa_{2}+s\left[c_{l,2}\ln\left(R\right)+\mathcal{O}(R^{0})\right]\:,\label{eq:Tau2nonshort}
\end{equation}
with $c_{l,2}=-\tilde{\beta}\frac{\pi^{2}}{6}$. Again, the term
scaling with $A/R$ is absent but a new term scaling with $\ln\left(R\right)$
appears. For spherical walls next order term is the radius independent
coefficient $\frac{1}{36}\pi^{2}\tilde{\beta}[\gamma_{\textrm{E}}-6-6\log(2)+\log(\tilde{\beta})]$
and term of order $R^{-1}$ is null.

The adopted approach to evaluate $\tau_{2}$ is easily extended to
systems with dimension $d\neq3$, which are also frequently studied.
For example, for $d=2$ the virial series equation of state of the
soft-disks system in bulk\citep{Briano_1981} has been previously
evaluated. In the case of a planar wall that cut the $d$-space in
two equal regions (one of which is available for particles), one should
replace in Eq. (\ref{eq:Cx}) $m$ by $d-1+m'$, $m'=0$ corresponds
to the bulk $b_{2}$ and $m'=1$ corresponds to the planar term $a_{2}$.
For a $d$-spherical wall one finds that term of order $R^{d-2}$
($m'=2$) is zero and $m'=3$ corresponds to $c_{2}$ (order $R^{d-3}$).
Expressions of $S(u,r)$ for $d\neq3$ were given in Ref. \citep{Urrutia_2010}.
\begin{figure}
\begin{centering}
\includegraphics[width=0.85\columnwidth]{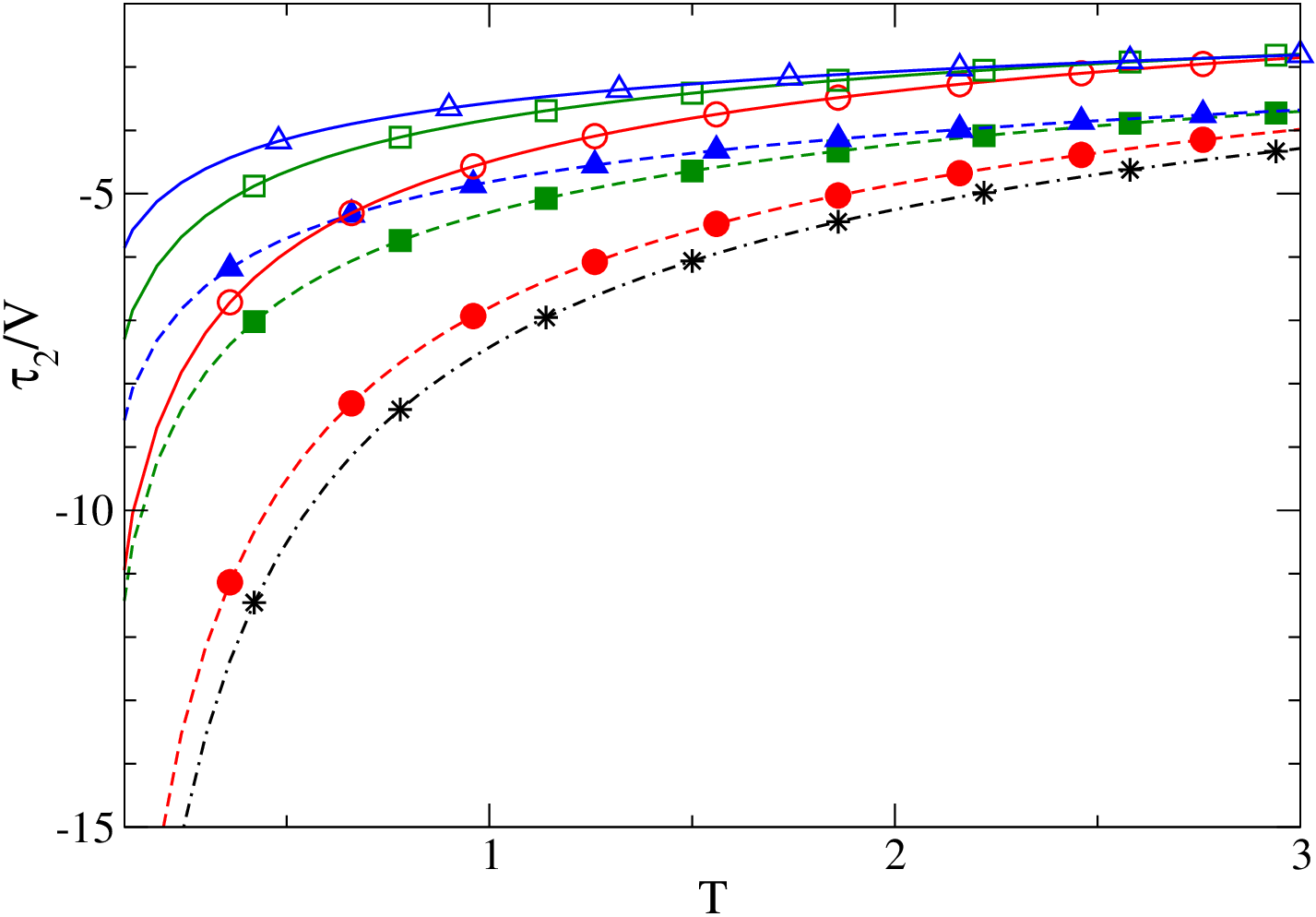}
\par\end{centering}

\caption{Second cluster integral divided by the volume for a fluid confined
in a spherical pore. Different values of $\nu$ corresponds to circles
($\nu=6$), squares ($\nu=9$) and triangles ($\nu=12$). Continuous
lines show results for $R=2$, dashed lines are for $R=20$ and dot-dashed
is for the bulk system (only $\nu=6$ is shown).\label{fig:Tau2RT}}
\end{figure}
As an example of the obtained results in Fig. \ref{fig:Tau2RT} we
plot the dependence with $T$ of the second cluster integral for the
soft-sphere IPL fluid confined in a spherical pore. Curves show different
values of the exponent and of the cavity radius. In the plot the natural
units for $T$ were used i.e. $T$ is measured in $\alpha/k_{\textrm{B}}$
units.

\section{Results: Surface tension\label{sec:Results}}

We consider the open system at low density confined by planar, spherical
or cylindrical walls and truncate Eq. (\ref{eq:Omgz}) at second order
to obtain $\beta\Omega=-zV-z^{2}\frac{1}{2}\tau_{2}$. Therefore the
first consequence of our calculus on $\tau_{2}$ is that the grand-free
energy of the system contains the expected terms linear with volume
and surface area. These terms are identical for the three studied
geometries. At planar geometry, no extra term exist as symmetry implies
for all $\tau_{i}$. In the case of spherical confinement a term linear
with total normal curvature of the surface $2A/R\propto R$ does not
appear at order $z^{2}$ but it should exist at higher ones. A term
linear with quadratic curvature $A/R^{2}\propto\textrm{constant}$
exists. Extra terms that scale with negative powers of $R$ were also
found. A logarithmic term proportional to $\ln R$ was recognized
only for $\nu=6$. The cylindrical confinement is similar to the spherical
case, thus we simply trace the differences: even that Gaussian curvature
is zero in this geometry, a term linear with $A/R^{2}\propto L/R$
was found. The existence of a logarithmic term for $\nu=6$ was verified,
in this case it was proportional to $L\ln R/R$.

For bulk homogeneous system the pressure and number density are $\beta P=z+z^{2}b_{2}$
and $\rho_{\textrm{b}}=z+z^{2}2b_{2}$ (subscript b refers to the
bulk at the same $T$ and $\mu$). On the other hand, the surface
tension is\citep{Urrutia_2014} 
\begin{equation}
\beta\gamma=-\frac{\Delta\tau_{2}}{2A}z^{2}=-\frac{\Delta\tau_{2}}{2A}\rho_{\textrm{b}}^{2}\:,\label{eq:Gamrhob}
\end{equation}
that are exact up to $\mathcal{O}(z^{3})$ and $\mathcal{O}(\rho_{\textrm{b}}^{3})$.
By collecting results from Eqs. (\ref{eq:Tau2},\ref{eq:Tau2plan},\ref{eq:Tau2sphe})
and (\ref{eq:Tau2short}), and replacing in (\ref{eq:Gamrhob}) one obtains the exact
expression for planar and spherical walls and an approximated expression for
cylindrical walls, up to the mentioned order in density. It yields
$\gamma=a_{2}T\rho_{\textrm{b}}^{2}$ for the planar case. When lower
order terms in $R^{-1}$ are retained for curved walls it is found,
\begin{equation}
\gamma_{\textrm{s}}=\left[a_{2}-\frac{c_{2}}{4\pi R^{2}}-\frac{d_{2}}{4\pi R^{3}}+\mathcal{O}\left(R^{-4}\right)\right]T\rho_{\textrm{b}}^{2}\:,\label{eq:gammaps}
\end{equation}
\begin{equation}
\gamma_{\textrm{c}}=\left[a_{2}-\frac{3c_{2}}{32\pi R^{2}}+\mathcal{O}\left(R^{-3}\right)\right]T\rho_{\textrm{b}}^{2}\:.\label{eq:gammac}
\end{equation}
For the special case $\nu=6$ we should replace $c_{2}$ with $c_{l,2}\ln R$
($d_{2}\neq0$ only if $\nu=7$).
\begin{figure}
\centering{}\includegraphics[width=0.85\columnwidth]{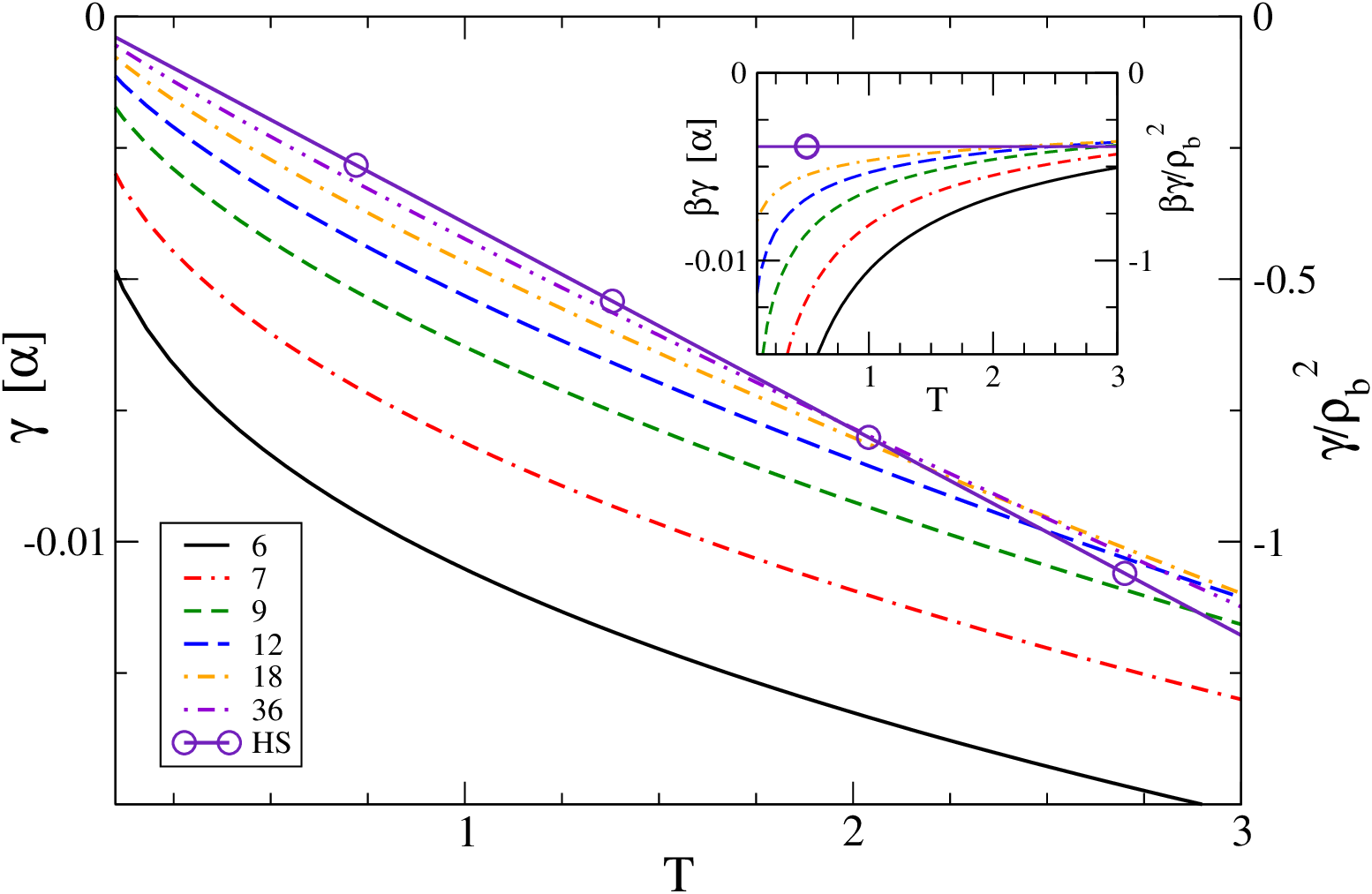}\caption{Fluid/wall
surface tension in the case of a planar wall; we fix $\rho_{\textrm{b}}=0.1$
and consider various $\nu$ values. From bottom to top (at low temperatures)
$\nu$ increases. Curves correspond to $\nu=6,7,9,12,18,36$ and to
HS ($\nu\rightarrow\infty$).\label{fig:gammaPlanar}}
\end{figure}
In Fig.\ref{fig:gammaPlanar} it is shown the surface tension of the
gas confined by a planar wall for different values of $\nu$. Scale
on the right shows $\gamma/\rho_{\textrm{b}}^{2}$, which is independent
of density. All cases show $\gamma<0$, which is consistent with a
repulsive potential and a monotonous decreasing behavior of $\gamma$
with $T$. In the limit $\nu\rightarrow\infty$ we obtain the asymptotic
curve, which is a straight-line in coincidence with the HS result.
In the inset it is shown $\beta\gamma$. There, asymptotic behavior
for large $T$ corresponds to the constant value HS result and the
hardening of curves with increasing $\nu$ is apparent. We note that
several curves cross the HS limiting line and also that lines of different
hardness intersect. This shows that softer potential may produce both
smaller surface tension than harder potentials (at low temperature)
but also larger surface tension than harder potentials (at high temperature).
In Table \ref{tab:gamk} we present the dependence of $\gamma$ with
temperature for planar walls.
\begin{table}
\centering{}%
\begin{tabular}{|c|c|c|c|}
\hline 
$\nu$ & $-\gamma/\rho_{\textrm{b}}^{2}$ & $k/\rho_{\textrm{b}}^{2}$ & $-k/\gamma$\tabularnewline
\hline 
\hline 
$6$ & $1.052T^{1/3}$ & $0.0982$ & $0.09$\tabularnewline
\hline 
$7$ & $0.812T^{3/7}$ & $0.107T^{1/7}$ & $0.13$\tabularnewline
\hline 
$8$ & $0.696\sqrt{T}$ & $0.0593T^{1/4}$ & $0.08$\tabularnewline
\hline 
$9$ & $0.629T^{5/9}$ & $0.0438T^{1/3}$ & $0.07$\tabularnewline
\hline 
$12$ & $0.532T^{2/3}$ & $0.0290\sqrt{T}$ & $0.053$\tabularnewline
\hline 
$18$ & $0.467T^{7/9}$ & $0.0221T^{2/3}$ & $0.047$\tabularnewline
\hline 
$36$ & $0.423T^{8/9}$ & $0.0185T^{5/6}$ & $0.042$\tabularnewline
\hline 
$\infty$ & $0.393T$ & $0.0164T$ & $0.042$\tabularnewline
\hline 
\end{tabular}\caption{Dependence of planar surface tension and bending rigidity with temperature
for some $\nu$ values. We fix $\alpha=k_{\textrm{B}}=1$. The ratio
$-k/\gamma$ was evaluated at $T=1$.\label{tab:gamk}}
\end{table}

\begin{figure}
\centering{}\includegraphics[clip,width=0.85\columnwidth]{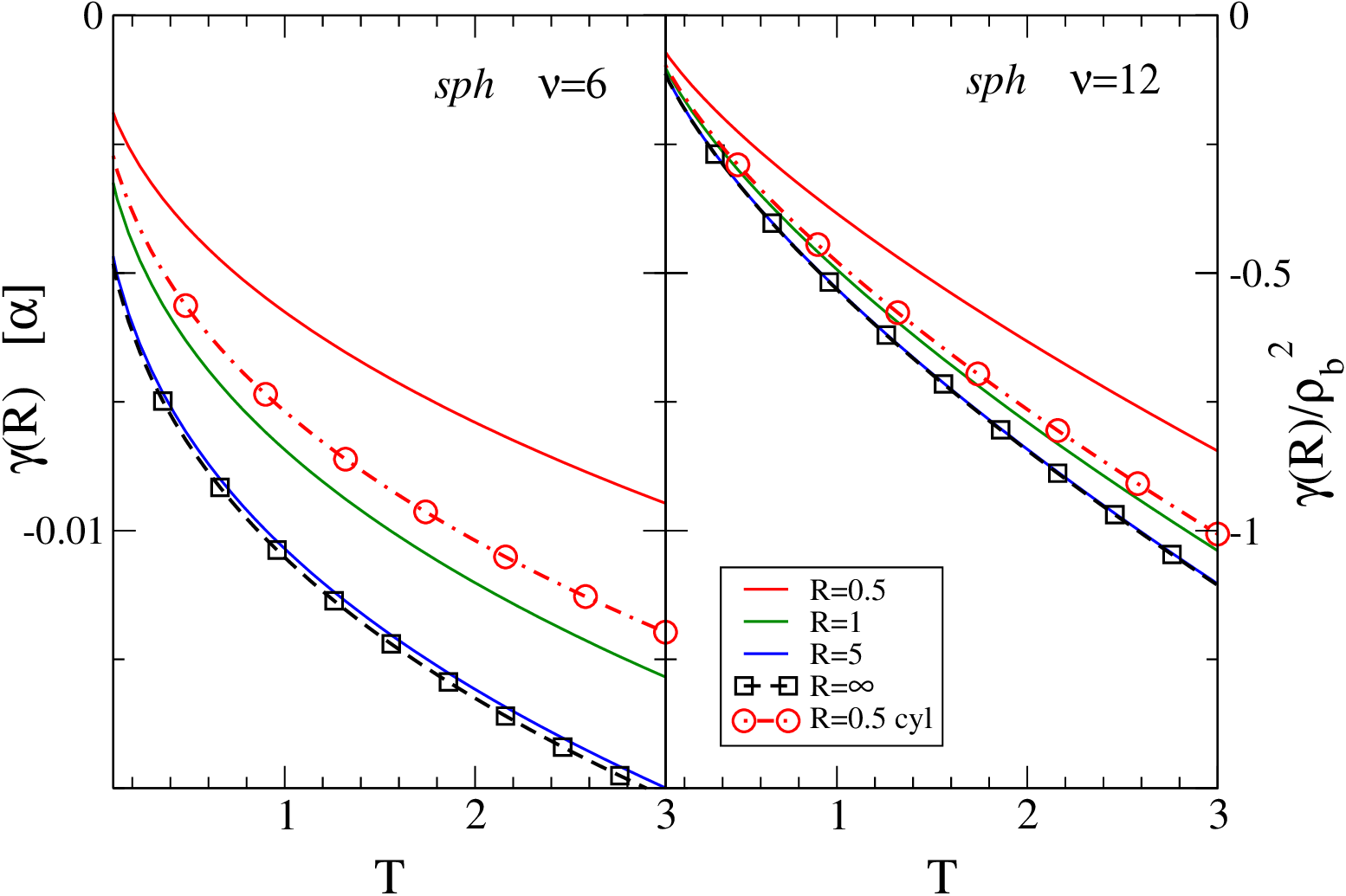}\caption{Surface tension of the fluid confined by spherical walls at $\rho_{\textrm{b}}=0.1$
(for both concave and convex shapes) and at various radii. At the
left we plot the case $\nu=6$, at right the case $\nu=12$. The planar
limit and cylindrical cases are also shown for comparison.\label{fig:gammaCurv}}
\end{figure}
 In the case of spherical walls the curvature dependence of the surface
tension is plotted in Fig. \ref{fig:gammaCurv}. There, results for
the $\nu=6$ (softer) and $\nu=12$ (harder) systems as a function
of temperature are shown for different values of $R$. Again we found
that surface tension is negative and decreases with $T$, which are
characteristic signatures of repulsive interactions. Surface tension
becomes larger at smaller radius and at $R\gtrsim5$ is well described
by the planar wall limit. A comparison of cases $\nu=6$ and $\nu=12$
shows that the sensitiveness of $\gamma$ with the radius is larger
at softer potential.

Figure \ref{fig:gammaCurv} is also related with the excess surface
adsorption $\Gamma_{A}=\left(\left\langle n\right\rangle -\left\langle n\right\rangle _{\textrm{b}}\right)/A$.
Series expansion of $\Gamma_{A}$ up to order $z^{2}$ and $\rho_{\textrm{b}}^{2}$
are: $\Gamma_{A}=z^{2}\Delta\tau_{2}/A$$=\rho_{\textrm{b}}^{2}\Delta\tau_{2}/A$.
Thus, up to the order of Eq. (\ref{eq:gammaps}) it is 
\begin{equation}
\Gamma_{A}=-2\gamma/T\:,\label{eq:Gaga}
\end{equation}
showing that curves of $\gamma(R)$ also plot $-\Gamma_{A}T/2$. Naturally,
the same apply to the planar case shown in Fig. \ref{fig:gammaPlanar}
and to the cylindrical one. It must be noted that $\gamma(R)$ and
$\Gamma_{A}$ depend on the adopted surface of tension that we fixed
at $r=R$ where external potential goes from zero to infinity. This
fixes the adopted reference region characterized by measures $V$,
$A$, and $R$. The effect of introducing a different reference region
on $\gamma(R)$ was systematically studied in Refs. \citep{Urrutia_2014,Urrutia_2015,Reindl_2015}
and will be briefly discussed in Sec. \ref{sec:Rigidities}.

\section{Results: Bending and Gaussian rigidities\label{sec:Rigidities}}

On the basis of our results the expansion given in Eqs. (\ref{eq:gammaks},
\ref{eq:gammakc}) is adequate for $\nu>6$ but not if $\nu=6$. For
$\nu>6$, we found $\gamma\delta=\mathcal{O}(\rho_{\textrm{b}}^{3})$,
\begin{eqnarray}
k & = & \frac{\pi}{192}\Gamma\left(1-\frac{6}{\nu}\right)\,T^{1-\frac{6}{\nu}}\,\rho_{\textrm{b}}^{2}+\mathcal{O}(\rho_{\textrm{b}}^{3})\:,\label{eq:kkbar}\\
\bar{k} & =- & \frac{\pi}{288}\Gamma\left(1-\frac{6}{\nu}\right)\,T^{1-\frac{6}{\nu}}\,\rho_{\textrm{b}}^{2}+\mathcal{O}(\rho_{\textrm{b}}^{3})\:.
\end{eqnarray}
Again, if we replace using the identity $\beta^{q}\Gamma\left(1-q\right)\rightarrow-qC_{q}(0)$
these expressions coincide with those found recently for the Lennard-Jones
fluid, but with a different definition for $C_{q}(0)$.\citep{Urrutia_2016}
\begin{figure}
\centering{}\includegraphics[width=0.85\columnwidth]{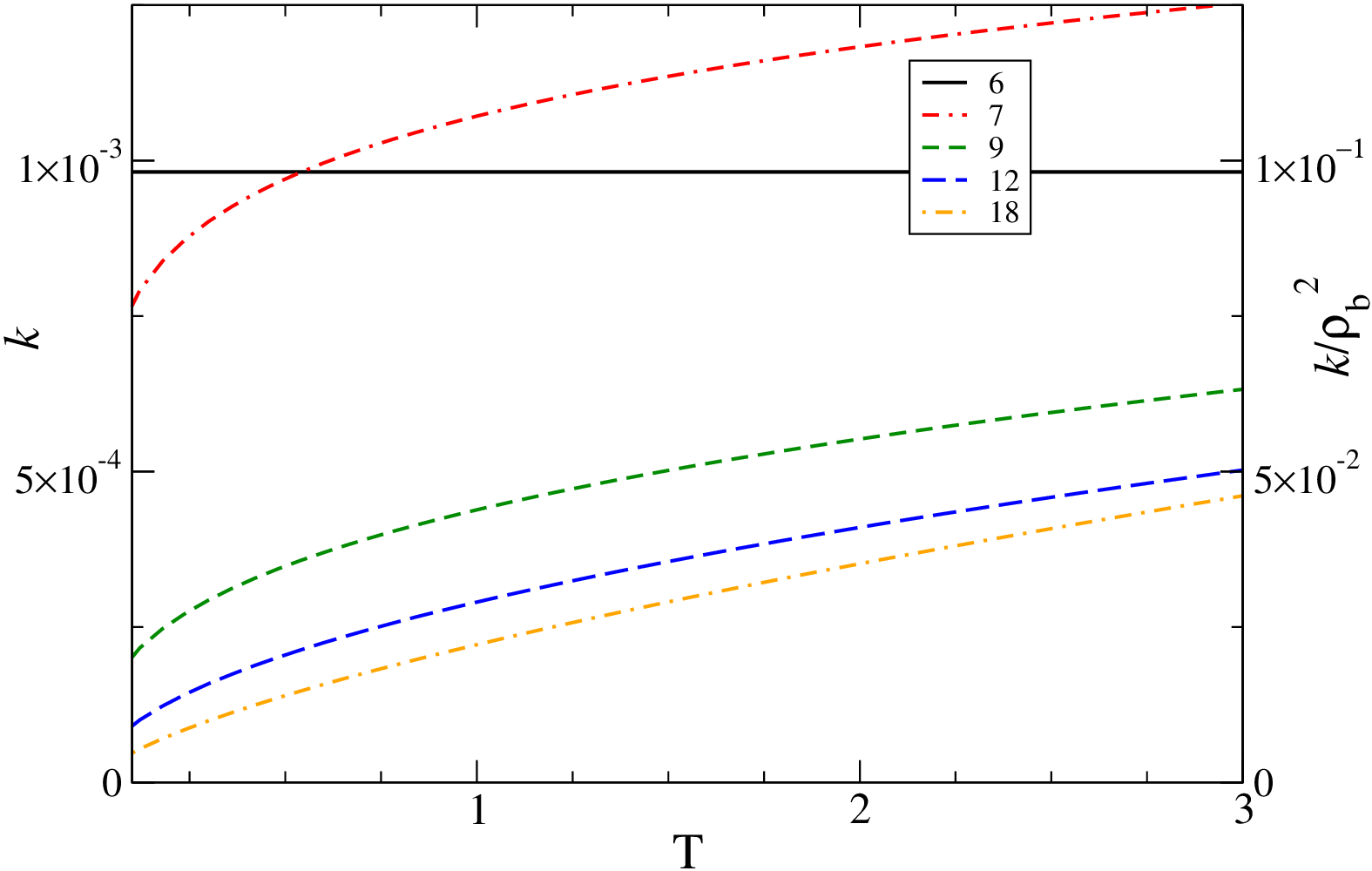}\caption{Fluid/wall bending rigidity $k$ as a function of temperature. We
fix $\rho_{\textrm{b}}=0.1$ and consider various $\nu$ values.
Curves correspond to $\nu=6,7,9,12,18$. The Gaussian rigidity $\bar{k}$
was not plotted because $\bar{k}=-0.66\boldsymbol{k}$ {[}see Eq. (\ref{eq:qkk}){]}.\label{fig:k}}
\end{figure}
In Fig. \ref{fig:k} the bending rigidity constant $k$ is presented
as a function of temperature for different values of hardness parameter
$\nu$. It is a positive increasing function of $T$ and is smaller
for higher $\nu$. The case $\nu=6$ is different because the $\ln R$
term. Gaussian rigidity $\bar{k}$ is a negative decreasing function
of $T$ and is higher for higher $\nu$. In Table \ref{tab:gamk}
we present the numerical coefficients of the bending rigidity to show
order of magnitude of $k\left(T\right)$. Besides, the relative weight
of $k$ in surface tension is shown in last column. We observe that
$k$ is smaller than $\gamma$ but may be as large as $0.13\times\gamma$
(case $\nu=7$ and $T=1$). The order $R^{-3}$ term in Eq. (\ref{eq:gammaks})
corresponds to $\mathscr{C}=-\frac{\pi}{96}\rho_{\textrm{b}}^{2}$
for $\nu=7$ and is zero otherwise. It is interesting to calculate
the quotient between $k$ and $\bar{k}$, and also the quotient of
the next to $R^{-1}$ term in $\gamma$ between spherical and cylindrical
cases. For all $\nu>6$ one finds 
\begin{eqnarray}
k/\bar{k} & = & -3/2\:,\label{eq:qkk}\\
2\frac{2k+\bar{k}}{k} & = & 8/3\:.\label{eq:2kpkbs}
\end{eqnarray}
Remarkably, they are universal values in the sense that are independent
of both $\nu$ and the state variable $T$. In the last ratio, the
left-hand side of equation is independent of the assumptions of a
Helfrich-based expression for $\gamma(R)$, and therefore it still applies
if Eqs. (\ref{eq:gammaks},\ref{eq:gammakc}) were wrong.

For non-short-ranged interactions as in the case of $\nu=6$ the logarithmic
term makes Helfrich expansion\citep{Helfrich_1973} of $\gamma(R)$
in power of $R^{-1}$ no longer valid. Thus, for $\nu=6$ instead
of the Eqs. (\ref{eq:gammaks},\ref{eq:gammakc}), one obtains for
the spherical and cylindrical walls 
\begin{eqnarray}
\gamma_{\textrm{s}}(R) & = & \gamma-\frac{2\gamma\delta}{R}+\left(2k+\bar{k}\right)\frac{\ln R}{R^{2}}+\mathcal{O}(R^{-2})\:,\label{eq:gammaksLR}\\
\gamma_{\textrm{c}}(R) & = & \gamma-\frac{\gamma\delta}{R}+k\frac{\ln R}{2R^{2}}+\mathcal{O}(R^{-2})\:,\label{eq:gammakcLR}
\end{eqnarray}
where bending and Gaussian rigidities were identified with the next
order terms beyond $\gamma\delta$. We found 
\begin{equation}
k=\frac{\pi}{32}\rho_{\textrm{b}}^{2}+\mathcal{O}(\rho_{\textrm{b}}^{3})\:\:\:\textrm{ and }\:\:\:\bar{k}=-\frac{\pi}{48}\rho_{\textrm{b}}^{2}+\mathcal{O}(\rho_{\textrm{b}}^{3})\:\:.\label{eq:kkbarLR}
\end{equation}
In this case both rigidities are temperature independent. The advent
of $\ln R$ terms in Eqs. (\ref{eq:gammaksLR},\ref{eq:gammakcLR})
demand to revise the invariance under the change of reference. $A\gamma_{\textrm{s}}(R)$
produces in $\Omega$ a term $\left(2k+\bar{k}\right)\ln R$, which
is invariant and $A\gamma_{\textrm{c}}(R)$ produces in $\Omega$
a term $k\ln R/R$, which is also invariant. Both terms are invariant
under the change of reference. Thus, for $\nu=6$ both rigidities
$k$ and $\bar{k}$ are invariant under the change of reference system. 

Even for $\nu=6$ we find for the ratios of curvatures the universal
results given in Eqs. (\ref{eq:qkk},\ref{eq:2kpkbs}). In fact, the
origin of these fundamental values is purely \emph{geometrical} and
was obtained previously for HS, square well, and Lennard-Jones, potentials.\citep{Urrutia_2014,Urrutia_2016}
Thus, essentially any pair interaction potential between particles
produce the same value for the ratio $k/\bar{k}$ at low density.
This result is in line with that found numerically using a second-virial
approximation DFT.\citep{Reindl_2015} The same \emph{geometrical}
status claimed for $k/\bar{k}$ corresponds to the result $\gamma\delta=0+\mathcal{O}(\rho_{\textrm{b}}^{3})$
that is directly derivable from Eqs. (\ref{eq:Tau2plan},\ref{eq:Tau2sphe},\ref{eq:Tau2cyl})
and applies to essentially any pair potential.

\subsection*{Accuracy of truncation in the low curvature expansion}

Based on the exact universal relation Eq. (\ref{eq:qkk}) we analyze
the consequences of truncate higher order curvature terms in $\gamma(R)$
and discuss some particular aspects concerning soft-spheres. We drop
terms beyond $k$ and $\bar{k}$ and use Eq. (\ref{eq:qkk}) to rewrite
surface tension as a function of only one rigidity constant, e.g.,
$\bar{k}$,
\begin{eqnarray}
\gamma_{\textrm{s}}(R) & = & \gamma-\frac{2\gamma\delta}{R}-2\frac{\bar{k}}{R^{2}}\ell\:,\label{eq:gammaks-U}\\
\gamma_{\textrm{c}}(R) & = & \gamma-\frac{\gamma\delta}{R}-\frac{3\bar{k}}{4R^{2}}\ell\:,\label{eq:gammakc-U}
\end{eqnarray}
where $\ell=1$ or $\ell=\ln R$ as appropriate (e.g., for IPL if $\nu>6$
then $\ell=1$ and if $\nu=6$ then $\ell=\ln R$). Now, we look for
a simple relation that linking the properties of a fluid in a spherical
and cylindrical confinement (the same fluid under the same thermodynamic
conditions $T$, $\mu$) enables to measure accurately intrinsic curvature-related
properties. We focus on that producing the same surface tension, 
\begin{equation}
\gamma_{\textrm{s}}(R_{\textrm{s}})=\gamma_{\textrm{c}}(R_{\textrm{c}})\:;\label{eq:isogamma}
\end{equation}
i.e., for a spherical cavity with a given radius $R_{\textrm{s}}$
we obtain the radius of the cylindrical cavity producing the same
surface tension. Following Eqs. (\ref{eq:gammaks-U},\ref{eq:gammakc-U})
and including terms of $\mathcal{O}\left(\rho_{b}^{2}\right)$, this
surface isotension condition gives
\begin{equation}
R_{\textrm{c}}^{2}=0.375\,R_{\textrm{s}}^{2}\:,\label{eq:RcRs}
\end{equation}
for $\ell=1$ (the case $\ell=\ln R$ does not yield a simple analytic
result).
\begin{figure}
\begin{centering}
\includegraphics[width=0.85\columnwidth]{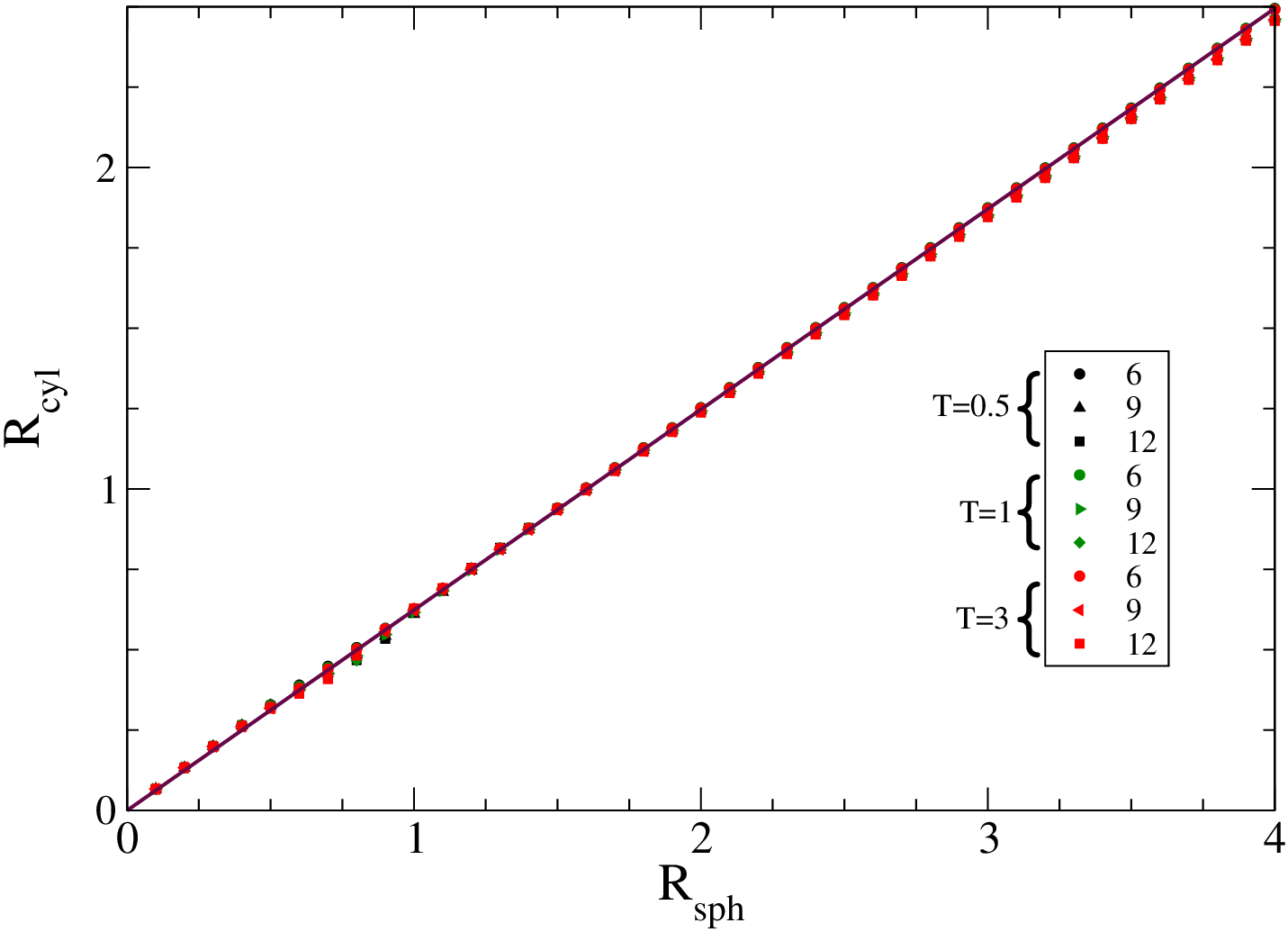}
\par\end{centering}

\caption{Relation between the radii under the iso-tension condition. Temperatures
$T=0.5,1,3$ are drawn in black, green, and red symbols, respectively,
but are difficult to distinguish in the plot. The straight line corresponds
to Eq. (\ref{eq:RcRs}).\label{fig:RcRs}}
\end{figure}
This is a remarkable simple relation. To test the accuracy of the
$R_{\textrm{c}}\leftrightarrow R_{\textrm{s}}$ relation beyond the
truncation of higher-order terms in Eqs. (\ref{eq:gammaks-U},\ref{eq:gammakc-U})
we solved numerically Eq. (\ref{eq:isogamma}) with the \emph{exact} 
$\gamma_{\textrm{s}}(R_{\textrm{s}})$ and a high-order truncation
for $\gamma_{\textrm{c}}(R_{\textrm{c}})$ (we include contribution
up to $C_{10/\nu}$). In Fig. \ref{fig:RcRs} are shown the obtained
results for the iso-tension relation between the radii of cylindrical
and spherical confinements for different hardness parameter $\nu$
and temperatures. The plot shows that linear behavior predicted by
Eq. (\ref{eq:RcRs}) is very robust applying for all $\nu\geq6$ and
for a broad range of temperatures and radii. It also checks the robustness
of approximate Eqs. (\ref{eq:gammaks-U},\ref{eq:gammakc-U}) that
would be good approximations for any fluid at low density.

Using Eq. (\ref{eq:Gaga}) we infer that the relation between $R_{\textrm{c}}$
and $R_{\textrm{s}}$ also apply to the surface isoadsorption $\Gamma_{\textrm{c}}(R_{\textrm{c}})=\Gamma_{\textrm{s}}(R_{\textrm{s}})$
condition. The isotension-isoadsorption relation Eq. (\ref{eq:RcRs})
is the consequence of purely \emph{geometrical} aspects and thus applies
to a large variety of fluids independently of the details of the interaction
potentials. At finite and small value of $\rho_{\textrm{b}}$ and
large enough radius the term $\gamma\delta\propto\rho_{\textrm{b}}^{3}$
should drive the relation between $R_{\textrm{c}}$ and $R_{\textrm{s}}$.
In such case the slope change according to $R_{\textrm{c}}^{2}=0.25\,R_{\textrm{s}}^{2}$.
This behavior is apparent in Ref. \citep{Reindl_2015} [see Fig. 4(a)
therein]. Through the measure of adsorption isotherms (using for example
molecular dynamics or Montecarlo simulations) we propose that relation
$\Gamma_{\textrm{c}}(R_{\textrm{c}})=\Gamma_{\textrm{s}}(R_{\textrm{s}})$
and Eq. (\ref{eq:RcRs}) are valuable tools to evaluate the accuracy
of different approximations and the importance of $\mathcal{O}\left(R^{-2}\right)$
terms in the curvature dependence of the adsorption and surface tension
for low-density fluids.

It is interesting to compare the relation $k=-3\bar{k}/2$ with that
used in the context of the morphometric approach, where the bending
rigidity identified with a quadratic term in the free energy is dropped.\citep{Reindl_2015}
To this end we use the same interface convention adopted above and
focus on low density behavior. The morphometric approach fix $k=0$
in Eqs. (\ref{eq:gammaks}) and (\ref{eq:gammakc}) for any density, giving
\begin{eqnarray}
\gamma_{\textrm{s}}(R) & \approx & \gamma-\frac{2\gamma\delta}{R}+\frac{\bar{k}}{R^{2}}\ell\:,\label{eq:gammaks-MRPH}\\
\gamma_{\textrm{c}}(R) & \approx & \gamma-\frac{\gamma\delta}{R}\:,\label{eq:gammakc-MRPH}
\end{eqnarray}
which must be compared with Eqs. (\ref{eq:gammaks-U}) and (\ref{eq:gammakc-U})
which are exact up to order $\ell R^{-2}$. Under the morphometric
approximation the correction to $\gamma_{\textrm{s}}$ produced by
the term $R^{-2}$ is \emph{opposite} in sign to the real one and
the inaccuracy introduced in the approximation of $\gamma_{\textrm{s}}$
has the same order of that introduced in $\gamma_{\textrm{c}}$. Then,
it is preferable to fix $k\approx0$ and $\bar{k}\approx0$ to obtain
both simpler expressions and more accurate results for $\gamma_{\textrm{s,c}}$
than those based on morphometric Eqs. (\ref{eq:gammaks-MRPH}) and (\ref{eq:gammakc-MRPH}).
Besides, at order $\rho_{\textrm{b}}^{2}$ morphometric approximation
yields that $\gamma_{\textrm{c}}(R_{\textrm{c}})=\gamma_{\textrm{s}}(R_{\textrm{s}})$
never happens which confirm its sensibility to high order curvature
terms.

As was mentioned the obtained results pertain to a reference surface
that coincides with the position of zero-to-infinite wall interaction.
The adopted reference surface has several advantages. For example,
for the ideal gas it gives the beautifully simple relation $\Omega=-PV$
and $\gamma=\gamma\delta=k=\bar{k}=0$ while for shifted surfaces
the free energy $\Omega$ becomes unnecessarily complicated. Further,
several \emph{universal} relations only apply under the adopted convention
as the low density behavior $\gamma=\mathcal{O}(\rho_{\textrm{b}}^{2})$,
$k=\mathcal{O}(\rho_{\textrm{b}}^{2})$ , $\bar{k}=\mathcal{O}(\rho_{\textrm{b}}^{2})$,
$\gamma\delta=\mathcal{O}(\rho_{\textrm{b}}^{3})$, the surface tension and
adsorption relation Eq. (\ref{eq:Gaga}), the rigidity constants ratios given
in Eqs. (\ref{eq:qkk}) and (\ref{eq:2kpkbs}), and isotension Eq. (\ref{eq:RcRs}).
Even more, it has been shown that the adopted reference provides the
more sensible condition to measure higher-order curvature terms in
free energy.\citep{Reindl_2015} Beyond these qualities, once the
properties are obtained on a given convention, one can transform to
different shifted surfaces by simple rules that linearly combines
$P$, $\gamma$, $\gamma\delta$, etc.\citep{Urrutia_2014}

\section{Summary and Conclusions\label{sec:Summary}}

The use of virial series for confined fluids is an unusual approach
that allows us to find new exact analytic results. This is a valuable
feature that contributes to develop the theoretical framework of inhomogeneous
fluids, a field where exact results are difficult to obtain and thus
scarce.

In this work we utilized virial series at the lowest nontrivial order
(up to order two in density and activity) to study the soft-sphere
system confined by hard walls of planar, spherical and cylindrical
shape. In the first and second cases we evaluate on exact grounds
the second cluster integral with its full dependence on $R$, $T$,
and $\nu$, while for cylindrical walls we found a quickly convergent
expansion. With these analytic expressions we systematically analyze
the effect of wall-curvature obtaining for the first time the expansion
for planar and curved wall-fluid surface tension and its curvature
components: Tolman length, bending and Gaussian rigidities. Even more,
we evaluated the next-to constant rigidity term for spherical confinement,
which is invariant under reference region transformation.

Our results for low density soft-spheres show that planar surface
tension is a negative and monotonously decreasing function in $T$,
as it is also the case for spherical and cylindrical walls. Furthermore,
the effect of softening-hardening of the IPL pair potential is non-monotonous:
for each $\nu$ there is a temperature where surface tension (and
surface adsorption) coincides with that of HS system, for smaller
temperatures $\gamma<\gamma_{HS}$ while for larger temperatures $\gamma>\gamma_{HS}$.
This inversion appears to be in the same direction of that found for
colloid-polymer mixtures where soft repulsion enhances the depletion
mechanism.\citep{Rovigatti_2015} For the dependence on curvature
it is observed that surface tension decreases with decreasing $R$
and that $\gamma<\gamma_{\textrm{c}}<\gamma_{\textrm{s}}$ at least
for radii as smaller as $R\approx0.5$. 

In the case of curved walls we analyzed the small curvature expansion
of surface tension and verify the existence of a logarithmic term
when $\nu=6$. We calculated the exact expressions of bending and
Gaussian rigidities as well as the simple relation between them. Bending
rigidity is a positive increasing function of $T$, which decreases
with rigidity $\nu>7$ but is constant if $\nu=6$. 

We verified the validity of a set of relations that apply to \emph{any}
low-density fluid confined by smooth walls. They involve surface tension,
surface adsorption, Tolman length, bending and Gaussian rigidities,
and radii of curvature. These universal relations were found by adopting
a particular choice of the reference region but concern to any interface
convention once the reference transformation is done. Specially interesting
was the surface isotension relation between $R_{\textrm{s}}$ and
$R_{\textrm{c}}$ that provides an accurate mechanism to identify
and measure high-order curvature dependence of surface tension. We
expect that future development of approximate theoretical tools for
confined fluids, including mixtures with macroscopic particles as
colloids, may be benefited from these results.

Based on the Hadwiger theorem has been proposed that bending rigidity
constant could be nearly zero\citep{Konig_2005} and thus would be
unnecessary to include it in the expansion of $\gamma(R)$. Using
the universal relations we show that the inaccuracy introduced by
truncation of the bending rigidity term in $\gamma(R)$ is the same
order of Gaussian rigidity term (at least for low density and hard
walls), and therefore is not well justified from the numerical standpoint.
Given that at least under the adopted interface convention the morphometric
approximation does not comply with universal relations it could be
better to ignore both rigidity constants than merely fix $k\approx0$.
In particular, for the soft-sphere system with $\nu=7$ the inaccuracy
in $\gamma(R=1)$ introduced by the morphometric approximation is
as large as $7\%$ (at $T=1$). Our results complement other recent
works, showing that $k\neq0$ for different fluids under different
circumstances and suggesting that morphometric thermodynamics has to
be used with caution.\citep{Reindl_2015,Urrutia_2016,Urrutia_2014,HansenGoos_2014,Blokhuis_2013c,Blokhuis_2013}

We think that arguments inducing to establish the
absence of nonlinear terms in the free energy of fluids in thermodynamics
and statistical mechanics should be revised at least when one recognizes
that almost any real (finite-size) fluid system is in some sense confined.
\begin{acknowledgments}
This work was supported by Argentina Grant No. CONICET PIP-112-2015-01-00417.
\end{acknowledgments}

%

\appendix

\section{Some properties of $C_{q}$ \label{Apsec:Gamma}}

We analyze $C_{q}$ at fixed $\tilde{\beta}$. When $\varepsilon\rightarrow+0$
for $0<q<1$ the function $C_{q}$ converges but it diverges for $q\geq1$.
In the convergent case we have 
\[
qC_{q}(0)=\tilde{\beta}^{q}q\Gamma\left(-q\right)=-\tilde{\beta}^{q}\Gamma\left(1-q\right)=-\tilde{\beta}C_{q-1}(0)
\]
an identity used to obtain Eqs. (\ref{eq:b2},\ref{eq:a2},\ref{eq:c2}),
while for the nonconvergent case one can transform through $q\Gamma(-q,\varepsilon)=-\Gamma(1-q,\varepsilon)+e^{-\varepsilon}\varepsilon^{-q}$
to obtain\citep{Abramowitz1972}
\begin{eqnarray}
qC_{q}(\varepsilon) & = & -\tilde{\beta}^{q}\Gamma(1-q,\tilde{\beta}\varepsilon)+\left(\tilde{\beta}\varepsilon\right)^{-q}\left(e^{-\tilde{\beta}\varepsilon}-1\right)\:,\label{apeq:qCqe}\\
 & = & -\tilde{\beta}C_{q-1}(\varepsilon)+\left(\tilde{\beta}\varepsilon\right)^{-q}\left(e^{-\tilde{\beta}\varepsilon}-1+\frac{\tilde{\beta}\varepsilon}{q-1}\right).\nonumber 
\end{eqnarray}

The functional behavior of $C_{q}$ is simpler to analyze by introducing
the function $F_{q}\equiv C_{q}(\varepsilon)\tilde{\beta}^{-q}$ that
depends on $z=\mathbf{\tilde{\beta}}\varepsilon$, but not on $\tilde{\beta}$
and $\varepsilon$ separately. The series expansion for small and
positive $z$ is 
\begin{equation}
F_{q}=\Gamma(-q)+z^{-q}\sum_{{\scriptstyle k=1}}^{\infty}\frac{(-z)^{k}}{(q-k)k!}\:,\label{eqap:Fser1}
\end{equation}
which applies to noninteger values $q>0$. On the other hand,
in the case of integer positive values of $q$,
\begin{equation}
F_{q}=\frac{(-1)^{q}}{q!}\left[H_{q}-\gamma_{E}+\ln z^{-1}\right]+z^{-q}\sum_{\stackrel{{\scriptstyle k=1}}{{\scriptstyle k\neq q}}}^{\infty}\frac{(-z)^{k}}{(q-k)k!}\:,\label{eqap:Fser2}
\end{equation}
where $\gamma_{\textrm{E}}$ is the Euler number and $H_{q}$ is the
harmonic number of order $q$ (for the lowest $q$ we have $H_{1}=1$,
$H_{2}=1.5$). Thus, $F_{q}\approx z^{1-q}$ for $q>1$ (and $q$
noninteger), but $F_{q}\approx\ln z^{-1}$ if $q=1$. Moreover, a
term proportional to $\ln z^{-1}$ appears for every integer value
$q\geq1$. On the opposite, for large values of $z>0$ we have the
following expansion:
\begin{eqnarray}
F_{q} & = & e^{-z}z^{-q}\left(\frac{1}{z}-\frac{q+1}{z^{2}}+\frac{(q+1)(q+2)}{z^{3}}\right.\nonumber \\
 &  & \left.-\frac{(q+1)(q+2)(q+3)}{z^{4}}+...\right)\:.\label{eq:FzL}
\end{eqnarray}
The asymptotic behavior for small (and positive) values of $q$ and
fixed $z$ is
\[
qF_{q}=-1+\mathcal{O}(q)\:,
\]
which reproduces the HS result.

\section{Hard core $C_{q}$\label{Apsec:HardCore}}

When the pair interaction between particles is defined as $\phi(r\leq\sigma)=+\infty$
and by the IPL given in Eq. (\ref{eq:POWpot}) for $r>\sigma$ we
obtain $C_{m+1,\nu}=-\frac{1}{m+1}+\frac{1}{\nu}C_{q}$, where the
first term on the right is the hard-core contribution and
\begin{eqnarray}
C_{q}(\varepsilon) & = & \int_{\varepsilon}^{1}y^{-(1+q)}\left[\exp\left(-\tilde{\beta}y\right)-1\right]dy\:,\label{eqap:CqHC}\\
 & = & \frac{1-\varepsilon^{-q}}{q}+\tilde{\beta}^{q}\left[\Gamma\left(-q,\tilde{\beta}\varepsilon\right)-\Gamma\left(-q,\tilde{\beta}\right)\right]\:.\nonumber 
\end{eqnarray}
This relation applies to both repulsive ($\tilde{\beta}>0$) and attractive
($\tilde{\beta}<0$) IPL potentials. In the last case it is convenient
to replace $\tilde{\beta}^{q}$ by $(-1)^{q}\left|\tilde{\beta}\right|^{q}$.
\end{document}